\documentclass[journal]{IEEEtran}

\usepackage{amsmath,graphics,amssymb,epsfig,subfigure,color,cite}

\usepackage{amsthm}
\theoremstyle{plain}
\newtheoremstyle{mystyle}
  {0mm}
  {0mm}
  {}
  {4mm}
  {\bfseries}
  {:}
  { }
  {\thmname{#1}\thmnumber{ #2}\thmnote{ (#3)}}
\theoremstyle{mystyle}

\usepackage{array}
\usepackage{multirow}
\usepackage{enumerate}
\usepackage{bm}
\usepackage{float}
\usepackage{makecell}
\usepackage[right=0.625in, left=0.625in, top=0.7in, bottom=1.1in]{geometry}

\usepackage{algorithm2e}
\usepackage{algpseudocode}
\usepackage{algorithmicx}
\usepackage{algcompatible}
\usepackage{setspace}
\usepackage{xcolor}
\definecolor{darkred}{RGB}{230,0,0}
\definecolor{darkgreen}{RGB}{0,130,0}
\definecolor{darkblue}{RGB}{0,0,200}

\usepackage{bbm}
\usepackage{bigints}
\usepackage{adjustbox}

\floatname{algorithm}{Algorithm}

\usepackage{enumitem}

\makeatletter
\newcommand{\vast}{\bBigg@{4.5}}
\newcommand{\Vast}{\bBigg@{7.5}}
\makeatother

\begin{document}
\title{\fontsize{22}{28}\selectfont
Context-Aware Wireless Token Communication \\via Joint Token Masking and Detection

}

\author{Junyong Shin, Joohyuk Park, Yongjeong Oh, Jihong Park, Jinho Choi, and Yo-Seb Jeon
	    \thanks{Junyong Shin, Joohyuk Park, and Yo-Seb Jeon are with the Department of Electrical Engineering, POSTECH, Pohang, Gyeongbuk 37673, Republic of Korea (e-mail: sjyong@postech.ac.kr; joohyuk.park@postech.ac.kr; yoseb.jeon@postech.ac.kr).}
        \thanks{Jinho Choi is with the School of Electrical and Mechanical Engineering, The University of Adelaide, SA 5005, Australia (email: jinho.choi@adelaide.edu.au).}
        \thanks{Jihong Park and Yongjeong Oh are with the Information Systems Technology and Design Pillar, Singapore University of Technology and Design, Singapore 487372 (email: jihong\_park, yongjeong\_oh@sutd.edu.sg).}
        \thanks{Yo-Seb Jeon and Jihong Park are corresponding authors.}
        }
	\vspace{-2mm}	
	
	\maketitle
	\vspace{-12mm}

\begin{abstract}
The increasing use of token-based representations in language-driven applications has motivated wireless token communication, where tokens are treated as fundamental units for transmission. However, conventional communication systems overlook dependencies among tokens and allocate transmission resources uniformly, leading to inefficient use of limited wireless resources under channel impairments.
In this paper, we propose a context-aware token communication framework that leverages a masked language model (MLM) as a shared contextual model between the transmitter (Tx) and receiver (Rx). At the Rx, we develop a context-aware token detection method that integrates channel likelihoods with MLM-based contextual priors under a Bayesian formulation, enabling robust token inference over noisy channels.
At the Tx, we propose a context-aware token masking strategy that selectively omits tokens that can be reliably inferred at the Rx, allowing the available power budget to be concentrated on more informative tokens.
These components are jointly designed through a shared MLM, establishing a unified Tx-Rx framework for efficient token transmission and detection.
Simulation results demonstrate that the proposed framework significantly improves reconstruction performance compared to conventional and existing token communication schemes, achieving up to 1.77× and 1.63× performance gains on the Europarl corpus and WikiText-103 datasets, respectively.
\end{abstract}


\section{Introduction}
Recent advances in natural-language processing have demonstrated the effectiveness of processing information through discrete tokens, which represent context-aware linguistic units. As wireless communications increasingly connects devices executing token-based applications, directly transmitting tokens over the air has emerged as a relevant and promising approach, which we refer to as {\em token communication} \cite{3GPP, TokComm_packet, TokComm_HLM, TokComm_crossmodal}.

Conventional communication systems are inherently context-agnostic, treating symbols as independent entities and overlooking the contextual relationships within token sequences. As a result, all tokens are transmitted with equal priority -even when some are easily predictable- and channel errors cannot be reliably resolved without contextual redundancy. Although recent studies have explored machine-learning-based symbol detection \cite{Detection_ContextLearning, Detection_Yujin}, these methods remain confined to minimizing local detection errors without leveraging inter-token dependencies. In contrast, token communication exploits contextual relationships among tokens, allowing systems to infer missing or corrupted tokens from surrounding context. As token-based applications continue to expand, this limitation highlights the need for a shift from context-agnostic to context-aware wireless communications.

Context modeling provides a principled way to capture dependencies among tokens and to characterize how information can be inferred from surrounding context. By quantifying token-level correlations within a sequence, context models enable both the identification of informative tokens for transmission and the recovery of missing or unreliable tokens. This capability is particularly important in wireless token communication, where limited transmission resources and channel impairments make it inefficient to treat all tokens equally, motivating context-aware transmission and detection strategies.

To effectively capture such contextual dependencies, masked language models (MLMs) provide a powerful framework for token communication systems. Specifically, trained to predict masked tokens from their surrounding context, MLMs capture bidirectional dependencies and provide a unified probabilistic representation of token sequences \cite{BERT, MLM}. Building on this capability, contextual priors derived from a shared MLM can be leveraged at both the transmitter (Tx) and receiver (Rx) in a unified manner. At the Rx, these priors can be integrated with channel observations under a Bayesian formulation to enable robust token detection. At the Tx, the same priors guide transmission decisions by identifying tokens that can be reliably inferred at the Rx, allowing the available power budget to be concentrated on more informative tokens. This leads to a joint Tx-Rx design in which context modeling directly governs both token detection and transmission strategies.

\subsection{Related Works}
Recent advances in learning-based communication have explored neural architectures that directly map source data to task-relevant outputs under channel impairments through end-to-end (E2E) optimization. A representative approach is deep joint source-channel coding (DeepJSCC), which jointly performs source compression and channel protection using a unified neural encoder-decoder architecture \cite{DeepJSCC, Semantic_Source_Channel, Semantic_Source_Channel2}. Several studies, such as \cite{Semantic_Source_Channel, Semantic_Source_Channel2}, have extended this framework to text transmission tasks. While such E2E frameworks have demonstrated strong task performance, they are typically trained under specific channel conditions or communication configurations. Consequently, adapting to heterogeneous wireless environments often requires training multiple models for different scenarios, leading to limited flexibility and substantial memory overhead, as studied in \cite{Blind, ESC-MVQ}.

To improve adaptability to varying communication environments, another line of research considers token-level transmission overhead control through masking strategies at the Tx. These approaches omit a subset of tokens during transmission and rely on a unified model to perform downstream tasks (e.g., classification) on partially observed token sequences.
For example, masked autoencoder (MAE)-based frameworks, such as \cite{MAE1}, randomly mask tokens according to a predefined ratio and perform downstream tasks using partially observed token sequences. While these approaches reduce transmission overhead, the masking process is random and ignores token importance. Moreover, the masking ratio is typically fixed or manually selected, limiting adaptability to both communication conditions and input instances.
To address this issue, attention-based token selection methods, such as \cite{attention_Jiwoong, attention_Joohyuk}, leverage attention scores \cite{attention} to estimate token importance and determine which tokens should be transmitted. In addition, recent approaches introduce learning-based token selection modules that are jointly trained with the underlying task model \cite{token_merging}. However, these methods are still tailored to specific tasks, making the resulting transmission strategy inherently task-dependent. As a result, their applicability to other tasks or integration with existing communication systems is limited, leading to reduced flexibility and compatibility.

To achieve task-independent operation, another line of research leverages {\em context modeling} to derive priors explicitly over token sequences. One representative approach incorporates these priors into a Bayesian framework for token detection at the Rx. In this framework, the Rx combines channel observations with prior probabilities modeled over token sequences. For instance, \cite{LLM-SC} employs pre-trained language models to provide contextual priors and performs autoregressive (AR) token detection by integrating these priors with channel likelihoods. While this approach incorporates contextual knowledge into token detection, the AR formulation captures only unidirectional token dependencies and cannot fully exploit bidirectional contextual relationships within the sequence. In addition, \cite{Semantic_Iter_Coding} proposes an iterative token recovery framework that refines token estimates by combining priors from the channel decoder with neural refinement modules. However, this approach requires additional neural decoders for each iterative refinement, which increases memory consumption and computational complexity and limits scalability in large-scale token communication systems. Moreover, the aforementioned approaches primarily exploit contextual information only at the Rx, leaving its potential use at the Tx largely unexplored.

To enable effective context modeling in token communication systems, MLMs can serve as a solution.
An MLM is a transformer-based model trained to predict masked tokens from their surrounding context, thereby learning the statistical dependencies within token sequences \cite{BERT}. MLMs model bidirectional dependencies by leveraging both preceding and succeeding tokens, enabling a unified probabilistic representation of token dependencies.
Building upon this context modeling capability, we consider a framework in which a shared MLM is used to provide contextual priors for both transmission and detection, enabling a joint Tx-Rx design for efficient token communication.

\begin{figure*}[t]
\centering
\includegraphics[width=1.85\columnwidth]{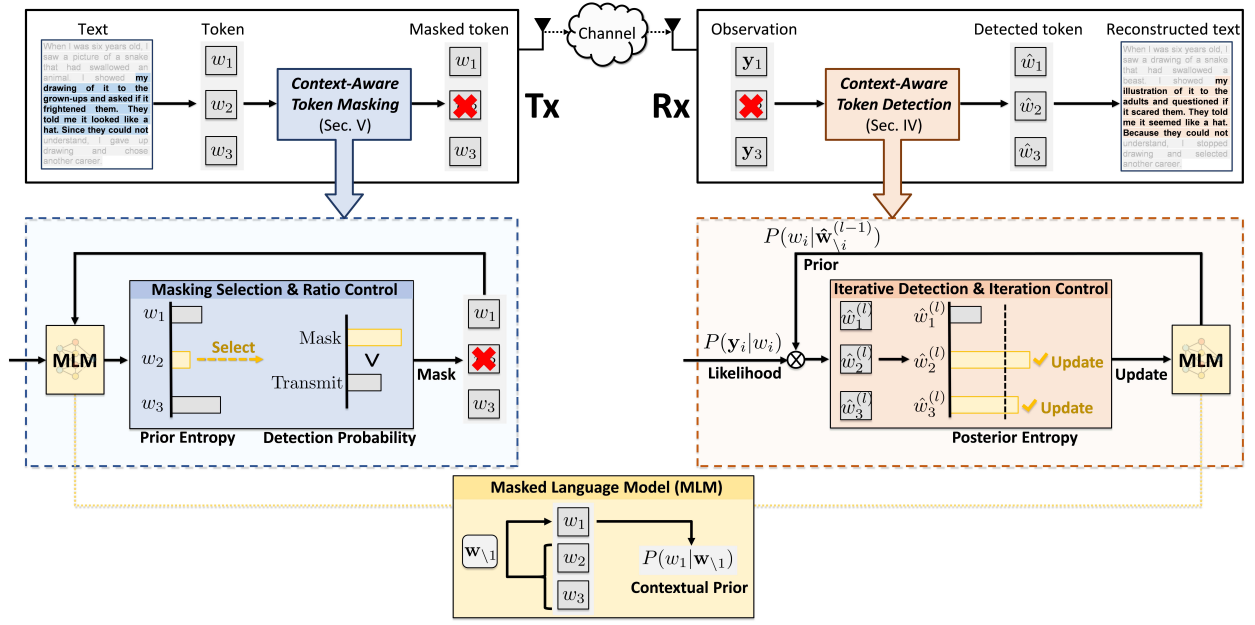}
\caption{An illustration of the proposed context-aware token communication framework.}
\label{Fig: overall illustration}
\vspace{-3mm}
\end{figure*}

\subsection{Contributions and Paper Organization}
The main contributions of this paper are outlined below:
\begin{itemize}
    \item We propose a context-aware token communication framework that employs a shared MLM to provide token-level contextual priors. This enables a unified context modeling of token priors without requiring additional task-specific architectures.
    \item We develop a context-aware token detection strategy at the Rx, which integrates channel likelihoods obtained from the observation with MLM-based contextual priors under a Bayesian formulation. The detection process is further enhanced by token-wise iteration control, which adaptively determines the required number of refinement iterations based on posterior uncertainty, enabling adaptive token inference under channel impairments.
    \item Building upon the Rx-side detection strategy, we propose a context-aware token masking strategy at the Tx. Specifically, an entropy-based sequential masking selection is employed to identify tokens that can be reliably inferred at the Rx, while a masking ratio control determines the appropriate number of tokens to omit by comparing token detection probabilities for masking decisions. This enables efficient allocation of the available power budget across tokens. Together with the Rx-side detection strategy, this forms a joint Tx–Rx design for context-aware token communication.
    \item Through extensive simulations, we show that the proposed framework significantly improves cosine similarity for text reconstruction, achieving up to 1.77× and 1.63× gains over conventional communication systems on the Europarl corpus and WikiText-103 datasets, respectively.

\end{itemize}

A conference version of this work was presented in \cite{AAAI} where we only introduce a context-aware token masking with a predefined masking ratio and iterative token detection with a fixed number of iterations.
To further enhance communication efficiency, in the current work, we newly develop adaptive masking ratio control at the Tx and token-wise iteration control at the Rx. These extensions enable dynamic adjustment and joint design of transmission and detection strategies based on channel conditions and prior uncertainty.

The remainder of this paper is organized as follows. Section~II presents the system model for wireless token communication. Section~III introduces MLM-based contextual prior modeling and an overview of the proposed joint Tx-Rx design. Section~IV describes the Rx-side context-aware token detection with token-wise iteration control. Section~V presents the proposed Tx-side context-aware masking with masking ratio control. Section~VI provides simulation results and performance evaluation. Finally, Section~VII concludes the paper.

\section{System Model}\label{Sec: System Model}

\subsection{Wireless Token Communication System}
We consider a point-to-point token communication scenario where a sequence of discrete language tokens is transmitted by a Tx over a wireless fading channel to its Rx. Let $\mathbf{w}$ denote the original token sequence of length $T$, given as
\begin{equation}
\mathbf{w}=[w_1,...,w_{T}],
\end{equation}
where $w_i$ represents the $i$-th token, which belongs to a vocabulary of size $V$ and can be represented by a binary vector of length $\lceil {\rm log}_2(V) \rceil$ through a tokenizer\footnote{ While we focus on tokens drawn from a finite vocabulary of size $V$, continuous token representations (e.g., latent embeddings) can be equivalently incorporated via discretization techniques such as vector quantization or 32-bit floating precision representations \cite{TokComm_crossmodal}.}.
Accordingly, each token element $w_{i}$ is converted into a bit sequence $\mathbf{b}_i$, defined as
\begin{equation}\label{eq: bit}
\mathbf{b}_i=[b_{i,1},...,b_{i,\lceil {\rm log}_2(V) \rceil}],\ \ \ b_{i,n}\in\{0,1\},\ \forall n.
\end{equation}
The bit sequence $\mathbf{b}_i$ is then grouped and modulated into complex-valued symbols.
The resulting symbol vector is defined as ${\sf s}(w_i) \in \mathcal{S}^K$, where each entry of the symbol vector represents a group of $m$ consecutive bits that are jointly mapped onto a complex symbol constellation $\mathcal{S}$, and $K\triangleq \left\lceil\lceil {\rm log}_2(V) \rceil/m\right\rceil$.
In our work, a $2^m$-ary quadrature amplitude modulation (QAM) scheme is employed for symbol mapping,
which is widely used in digital communication systems.

The modulated symbols are transmitted by the Tx over a Rayleigh block-fading channel \cite{Blockfading}. At the Rx, the received signal associated with the $i$-th token is expressed as
\begin{align}\label{eq: received}
    {\bf y}_i=h\sqrt\frac{P_{\rm tot}}{T}{\sf s}(w_i)+{\bf n}_i,
\end{align}
where $h\in \mathbb{C}$ denotes the complex fading coefficient that remains constant over one transmission block, $P_{\rm tot}$ is the total transmit power, and $\mathbf{n}_i\sim \mathcal{CN}(0,\sigma^2\mathbf{I})$ is additive white Gaussian noise (AWGN). 
The signal-to-noise ratio (SNR) at the Tx is given as
\begin{align}
    \text{SNR}=\frac{P_{\rm tot}\mathbb{E}[|h|^2]}{T\sigma^2}.
\end{align}

For a long sequence of source tokens, as illustrated in Fig.~\ref{Fig: overall illustration}, it can be segmented into multiple token sequences, each consisting of $T$ tokens, which are processed in parallel during transmission and detection.
Since the same token communication process is independently applied to each $T$-token sequence, the framework can be generalized to accommodate arbitrarily large input sources without altering its underlying operations.

\subsection{Context-Agnostic Token Communication and its Limitations} 
In conventional wireless communication systems, token detection is typically performed using the maximum likelihood (ML) detection \cite{MLdetection}, which selects the most likely token based on the received signal observation.
For the token system described above, the ML detection rule is given by
\begin{align}\label{eq: ML detection}
    \hat{w}_i^{{\rm ML}}=\underset{w_i}{{\rm argmax}}\ P(\mathbf{y}_i|w_i),
\end{align}
where $P({\bf y}_i|w_i)$ denotes the token-level likelihood function for $w_i$.
From \eqref{eq: received}, the function $P({\bf y}_i|w_i)$ is computed as
\begin{align}\label{eq: likelihood2}
    P(\mathbf{y}_i|w_i) 
    =\frac{1}{(\pi\sigma^2)^K}{\rm exp}\bigg(-\frac{1}{\sigma^2}\bigg\Vert{\bf y}_i-h\sqrt{\frac{P_{\rm tot}}{T}}{\sf s}(w_i)\bigg\Vert^2\bigg). 
\end{align}

At the Rx, the ML detection rule in \eqref{eq: ML detection} is optimal when each token $w_i$ is detected independently based solely on the corresponding observation ${\bf y}_i$. However, tokens within a given context dependent each other. 
Consequently, such \emph{context-agnostic} token detection, which ignotres the contextual information available at the Rx, does not maximize the overall token detection performance.

Similarity, at the Tx, conventional transmission strategies allocate resources (e.g., power) uniformly across tokens, without accounting for their contextual importance or predictability. Some tokens may not need to be transmitted if they can be reliably inferred from the context at the Rx, whereas others may carry critical information for understanding the overall context and thus require stronger protection. This suggests that such \emph{context-agnostic} token transmission can be suboptimal, particularly when the Rx exploits contextual information for token detection.

These limitations of context-agnostic operations at the Rx and the Tx motivate the development of a \emph{context-aware} token communication framework that explicitly incorporates contextual priors into both token detection and transmission. The proposed framework addresses these challenges through MLM-based contextual inference and joint Tx-Rx design, as detailed in the following sections.

\vspace{5mm}
\section{Overview of the Proposed Token Communication Framework}
This section presents an overview of the proposed token communication framework, which exploits contextual information enabled by a shared MLM. We first introduce contextual prior modeling based on the shared MLM (Sec.~\ref{subSec: contextual prior}). We then describe how these priors are incorporated into the Tx and Rx operations of the proposed framework (Sec.~\ref{subSec: overview}). The overall architecture is illustrated in Fig.~\ref{Fig: overall illustration}.

\subsection{Core Component: MLM as Contextual Token Prior Model}\label{subSec: contextual prior}
An MLM is well known for its ability to produce a conditional probability distribution over the vocabulary for a masked token position, by leveraging the surrounding tokens to infer the most plausible candidates \cite{BERT, MLM}. 
In the proposed token communication framework, the MLM is adopted as a shared contextual prior model at both the Tx and Rx.

To formalize the behavior of the MLM, let $[{\rm MASK}]$ denote the special mask token used by the MLM to mask a selected position. Given a token sequence $\mathbf{x} = [x_1,...,x_T]$ and a masking set $\mathcal{M}\subset\{1,...,T\}$, we define a masking operator ${\rm Mask}(\cdot)$ as
\begin{align}
    \mathbf{x}_{\rm m}&={\rm Mask}(\mathbf{x};\mathcal{M}),\\
    [\mathbf{x}_{\rm m}]_i&=
    \begin{cases}
        [{\rm MASK}],~i\in \mathcal{M},\\
        x_i, ~~~~~~~~~i \notin \mathcal{M}.
    \end{cases}
\end{align}
Based on this operator, the MLM models a conditional categorical distribution at each position as
\begin{align}\label{eq: mask BERT2}
    \hat{P}(x_i|\mathbf{x}_{\backslash i}) \triangleq[{\rm MLM}(\mathbf{x}_{\backslash i})]_i,
\end{align}
where $\mathbf{x}_{\backslash i}$ is the input sequence except the $i$-th token, defined as
\begin{align}\label{eq: x_minus i}
\mathbf{x}_{\backslash i} = [x_1,\ldots,x_{i-1},x_{i+1},\ldots, x_T].
\end{align}
Equivalently, for MLM inference, $\mathbf{x}_{\backslash i}$ can be represented by replacing the $i$-th token with $[{\rm MASK}]$ using the masking operator ${\rm Mask}(\cdot;\{i\})$:
\begin{align}
\mathbf{x}_{\backslash i} \equiv {\rm Mask}(\mathbf{x};\{i\}) = [x_1,\ldots,x_{i-1},[{\rm MASK}],x_{i+1},\ldots, x_T].
\end{align}

The conditional probability $\hat{P}(x_i|\mathbf{x}_{\backslash i})$ in \eqref{eq: mask BERT2} models the true distribution  $P(x_i|\mathbf{x}_{\backslash i})$, which provides a prior belief about $w_i$ given the unmasked context tokens.
We refer to this prior belief as the {\em contextual prior}, as it captures the token-level probability distribution conditioned on the surrounding tokens as context.
These token-wise contextual priors serve as the foundation for both token detection and masking  strategies.

\subsection{Joint Tx-Rx Strategy of the Proposed Framework}\label{subSec: overview}
The proposed framework is fundamentally based on a joint design of context-aware token detection at the Rx and context-aware token masking at the Tx, both guided by a shared contextual prior derived from the MLM. These two components are tightly coupled: the detection strategy enables the estimation of masked tokens by exploiting contextual priors inferred from unmasked tokens, while the transmission strategy leverages the same priors to selectively mask tokens that can be reliably reconstructed at the Rx.
The core components of each strategy are summarized below.

\subsubsection{Context-aware Token Detection Strategy at the Rx (Sec. IV)}
\begin{itemize}
\item {\bf Iterative token detection based on the MAP principle (Sec. IV-A):}
At the Rx, token detection is performed iteratively under the maximum a posteriori (MAP) principle by combining channel likelihood with MLM-based contextual priors.

\item {\bf Token-wise iteration control (Sec. IV-B):}
Since the reliability of token estimates varies across tokens, the number of refinement iterations $L_i$ is adaptively determined based on posterior uncertainty, avoiding unnecessary updates while maintaining reliable detection.

\end{itemize}

\subsubsection{Context-aware Token Masking Strategy at the Tx (Sec. V)} 
\begin{itemize}
    \item {\bf Sequential masking selection based on entropy scores (Sec. V-A):}
    The Tx exploits contextual priors from the shared MLM to identify tokens that can be reliably inferred without channel observations at the Rx, and sequentially masks highly predictable tokens using an entropy-based criterion.
    \item {\bf Masking ratio control (Sec. V-B):}
    To determine how many tokens to omit, the masking ratio $r$ is adaptively determined by comparing token detection probabilities for masking decisions at each step, and masking is terminated when further omission is no longer beneficial.

\end{itemize}

\vspace{10mm}
\section{Rx Strategy of the Proposed Framework: Context-Aware Token Detection}
This section presents the context-aware token detection method, corresponding to the Rx strategy of the proposed framework. The proposed detector iteratively refines token estimates by incorporating contextual priors generated by the shared MLM into the MAP framework.
We first describe iterative token detection based on the MAP principle (Sec.~IV-A), and then introduce an adaptive control mechanism that determines the required number of refinement iterations for each token based on its posterior entropy (Sec.~IV-B).

\subsection{Iterative Token Detection based on the MAP Principle}
We begin by describing the impact of token masking on the Rx-side signal model, noting that the masking strategy is applied at the Tx in the proposed framework. When token masking is employed, the Tx can allocate more power to the transmission of unmasked tokens, thereby improving their detection reliability.
Let $\mathcal{M} \subset \{1,\ldots,T\}$ denote the set of indices corresponding to the masked tokens, and let $N = |\mathcal{M}|$. Under a fixed total transmit power constraint $P_{\mathrm{tot}}$, the received signal for the unmasked tokens is then modified as follows:
\begin{align}\label{eq: received2}
    {\bf y}_i=h\sqrt\frac{P_{\rm tot}}{T-N}{\sf s}(w_i)+{\bf n}_i,\ \forall i\notin \mathcal{M}, %
\end{align}
and the token-level likelihood $P({\bf y}_i|w_i)$ for the unmasked token is expressed as
\begin{align}\label{eq: modified2}
    P(\mathbf{y}_i|w_i) 
    = \frac{1}{(\pi\sigma^2)^K}{\rm exp}\bigg(-\frac{1}{\sigma^2}\bigg\Vert{\bf y}_i-h\sqrt{\frac{P_{\rm tot}}{T-N}}{\sf s}(w_i)\bigg\Vert^2\bigg),
\end{align}
for all $i \notin\mathcal{M}$.

The ultimate goal of the proposed token detection is to exploit both the likelihood captured from the received signal and context-based priors that capture inter-token dependencies. To facilitate this, we consider the MAP token detection problem defined as
\begin{equation}\label{eq: MAP}
\hat{w}_{i}^{{\rm MAP}}=\underset{w_i}{{\rm argmax}}\ P(w_i|\mathbf{y}),
\end{equation}
where ${\bf y}=[\mathbf{y}_1^{\sf T},...,\mathbf{y}_{T}^{\sf T}]^{\sf T}$ is a total received vector associated with the token sequence ${\bf w}$.
Let $\mathbf{w}_{\backslash i}$ be the token sequence except the $i$-th token, defined as in \eqref{eq: x_minus i}.
Applying Bayes' rule and using the total probability theorem over $\mathbf{w}_{\backslash i}$ yields
\begin{align}\label{eq: pf1}
\hat{w}_i^{{\rm MAP}}&=\underset{w_i}{\text{argmax}}\ P(\mathbf{y}|w_i)P(w_i)\\
&=\underset{w_i}{\text{argmax}}\ \sum_{\mathbf{w}_{\backslash i}}P(\mathbf{y}|\mathbf{w})P(\mathbf{w}).
\end{align}
Moreover, using the token-conditional independence of observations, the MAP rule becomes
\begin{align}
&\hat{w}_i^{{\rm MAP}}=\underset{w_i}{\text{argmax}}\ \sum_{\mathbf{w}_{\backslash i}}P({\bf y}_i|w_i)P({\bf y}_{\backslash i}|{\bf w}_{\backslash i}) P(\mathbf{w})
\\&=\underset{w_i}{\text{argmax}}\ P({\bf y}_i|w_i) \sum_{\mathbf{w}_{\backslash i}}P({\bf y}_{\backslash i}|{\bf w}_{\backslash i})
  P(\mathbf{w})
\\&=\underset{w_i}{\text{argmax}}\ P({\bf y}_i|w_i) \sum_{\mathbf{w}_{\backslash i}} 
  P({\bf y}_{\backslash i}|{\bf w}_{\backslash i})
P(w_i|\mathbf{w}_{\backslash i})P(\mathbf{w}_{\backslash i}),\label{eq: before approx}
\end{align}
where ${\bf y}_{\backslash i} = [{\bf y}_1^{\sf T},\ldots,{\bf y}_{i-1}^{\sf T},{\bf y}_{i+1}^{\sf T},\ldots, {\bf y}_T^{\sf T}]^{\sf T}$.

Our formulation reveals that the MAP detection in \eqref{eq: before approx} accounts not only for the likelihood function $P({\bf y}_i|w_i)$ derived from the received signal ${\bf y}_i$, but also for the influence of other tokens in $\mathbf{w}_{\backslash i}$ when estimating $w_i$. This distinguishes it fundamentally from the conventional ML detection in \eqref{eq: ML detection}, which relies solely on ${\bf y}_i$ and ignores inter-token dependencies. Consequently, the MAP formulation establishes a theoretical basis for incorporating contextual information across tokens into the detection rule.

Despite the solid formulation in \eqref{eq: before approx}, directly solving the MAP detection problem requires marginalization over all possible combinations of $\mathbf{w}_{\backslash i}$, involving $V^{T-1}$ combinations.
Given the typical vocabulary size and sequence length considered in this work (e.g., $V=30522,T=128$), this computation is intractable.
To reduce complexity, we adopt a \emph{single-sequence approximation}, which has been widely used in iterative detection and decoding frameworks \cite{TurboEqualization}. Under this approximation, the prior distribution $P(\mathbf{w}_{\backslash i})$ is assumed to be concentrated on the most recently detected sequence $\hat{\mathbf{w}}_{\backslash i}^{(l-1)}$, i.e.,
\begin{align}\label{eq: single sequence}
    P(\mathbf{w}_{\backslash i})\approx
    \begin{cases}
        1,\ \ \ \text{if }\mathbf{w}_{\backslash i}=\mathbf{\hat w}_{\backslash i}^{(l-1)},\\
        0,\ \ \ \text{otherwise.}
    \end{cases}
\end{align}
Under this approximation, the summation in \eqref{eq: before approx} reduces to a single evaluation at $\hat{\mathbf{w}}^{(l-1)}$, which yields
\begin{align}\label{eq: iteration}
\hat{w}_i^{{\rm MAP}}\approx \underset{w_i}{\rm{argmax}}\ P({\bf y}_i|w_i)P(w_i|\mathbf{\hat w}_{\backslash i}^{(l-1)}).
\end{align}

The resultant context-aware token detection rule highlights a complementary relationship between two components: (i) the likelihood function $P({\bf y}_i|w_i)$, which quantifies the confidence in token $w_i$ based on the received signal ${\bf y}_i$, and (ii) the contextual prior $P(w_i|\mathbf{\hat w}_{\backslash i}^{(l-1)})$, which captures the probability of $w_i$ given the surrounding tokens.
When the SNR is sufficiently high, the likelihood term dominates the decision. In contrast, when inter-token dependencies are strong, the contextual prior becomes a more dominant role.
Based on the approximate MAP detection rule in \eqref{eq: iteration}, we devise a novel context-aware token detection rule where the contextual prior in \eqref{eq: iteration} is replaced with its practical approximation obtained from the MLM:
\begin{align}\label{eq: l-th prior}
    \hat{P}(w_i|\mathbf{\hat w}_{\backslash i}^{(l-1)})=[{\rm MLM}(\mathbf{\hat w}_{\backslash i}^{(l-1)})]_i,
\end{align}
as motivated by \eqref{eq: mask BERT2}.
At iteration $l \geq 1$, the token detection rule is given by
\begin{align}\label{eq: modified iteration}
    \hat{w}_i^{(l)}=
    \begin{cases}
        \underset{w_i}{\rm{argmax}}\ P({\bf y}_i|w_i)\hat{P}(w_i|\mathbf{\hat w}_{\backslash i}^{(l-1)}),\ &\text{if } i\notin \mathcal{M},\\
        \underset{w_i}{\rm{argmax}}\ \hat{P}(w_i|\mathbf{\hat w}_{\backslash i}^{(l-1)}),\ &\text{if } i\in \mathcal{M}.
    \end{cases}
\end{align}
The second case follows from the fact that $P({\bf y}_i|w_i)$ is treated as a constant for masked tokens, which are independent of $w_i$.

As discussed above, the intractability of the original MAP detection arises from the marginalization over all possible realizations of ${\bf w}_{\backslash i}$ in computing $P(w_i)$.
Instead, we adopt a modeling perspective in which the prior is represented through the dependence of $w_i$ on the surrounding tokens, i.e., as a contextual prior. In particular, under the approximation in \eqref{eq: single sequence}, this contextual prior is effectively characterized by the detected sequence from the previous iteration, enabling a tractable formulation that captures inter-token dependencies.
As the iteration index $l$ increases, the detected sequence becomes more reliable, leading to progressively refined contextual priors. This iterative refinement improves detection reliability and justifies the approximation adopted in \eqref{eq: single sequence}.

The iterative process of the proposed detection proceeds up to a predefined maximum number of MLM refinements, denoted by $L_{\rm max}$. At the initial iteration, no prior information is available at the Rx. Accordingly, the initial estimate $\mathbf{\hat w}_i^{(0)}$ $(l=0)$ is obtained based on conventional ML detection, i.e.,
\begin{align}
    \hat{w}_{i}^{(0)} = \begin{cases} 
        \underset{w_i}{\rm{argmax}}\ P({\bf y}_i|w_i), & \text{if}~ i\notin\mathcal{M},\\
        [{\rm MASK}], & \text{if}~ i\in\mathcal{M}.
    \end{cases}
\end{align}
In subsequent iterations ($1\leq l\leq L_{\rm max}$), the contextual prior is progressively refined using the updated token sequence based on \eqref{eq: modified iteration}.

\subsection{Token-Wise Iteration Control}
Although the maximum number of updates for each token is bounded by $L_{\rm max}$, different tokens generally require different numbers of refinement steps to reach reliable decisions. The major reason is that, depending on the received signal and the contextual prior, some tokens converge quickly, while others remain uncertain and require additional updates. To account for this variability, we first introduce a posterior entropy measure to quantify token-wise uncertainty. Based on this uncertainty measure, the proposed framework adaptively determines the required number of iterations for each token.

Entropy is widely used as a measure of uncertainty in probabilistic model outputs as well as in data detection \cite{attention_Jiwoong, Detection_ContextLearning}.
Our posterior entropy measure at the Rx is defined below.
\vspace{2mm}\\
{\bf Definition 1} (Posterior Entropy of Token Detection).
{\em At iteration $l$, the uncertainty of the $i$-th token is quantified by the posterior entropy, defined as}
\begin{align}\label{eq: adaptive iteration entropy}
    &H^{(l)}_{{\rm Rx},i}\nonumber\\
    &\triangleq\begin{cases}
    \!-\!\underset{v}{\sum}P^{(l)}\!({\bf y}_i|w_i\!=\!v,\mathbf{\hat w}_{\backslash i}^{(l-1)}) {\rm log}_2 P^{(l)}\!({\bf y}_i|w_i\!=\!v,\mathbf{\hat w}_{\backslash i}^{(l-1)}),\\~~~~~~~~~~~~~~~~~~~~~~~~~~~~~~~~~~~~~~~~~~~~~~~~~~~~\text{ if }i\notin\mathcal{M},\vspace{2mm}\\
    \!-\!\underset{v}{\sum}\hat{P}(w_i\!=\!v|\mathbf{\hat w}^{(l-1)}_{\backslash i})\text{log}_2\hat{P}(w_i\!=\!v|\mathbf{\hat w}^{(l-1)}_{\backslash i}),\text{ if }i\in\mathcal{M}.
    \end{cases}
\end{align}
Here, $P^{(l)}({\bf y}_i|w_i,\mathbf{\hat w}_{\backslash i}^{(l-1)})$ denotes the approximated posterior in \eqref{eq: iteration} at iteration $l$, given by
\begin{align}\label{eq: approx posterior}
    P^{(l)}({\bf y}_i|w_i,\mathbf{\hat w}_{\backslash i}^{(l-1)}) = \frac{1}{Z_i^{(l)}}P({\bf y}_i|w_i)\hat{P}(w_i|\mathbf{\hat w}_{\backslash i}^{(l-1)}),
\end{align}
 where $Z_i^{(l)}=\sum_v P({\bf y}_i|w_i=v)\hat{P}(w_i=v|\mathbf{\hat w}_{\backslash i}^{(l-1)})$ is the normalization factor of the $i$-th token at  iteration $l$. Furthermore, for masked tokens, the posterior entropy in \eqref{eq: adaptive iteration entropy} reduces to the prior entropy at the Rx.

For a predefined threshold $\eta$, if $H^{(l)}_{{\rm Rx},i}<\eta$, the $i$-th token is regarded as sufficiently reliable and excluded from further refinement; otherwise, it remains active.
Based on this uncertainty measure, the required number of MLM refinements for the $i$-th token is determined as
\begin{align}\label{eq: adaptive iteration1}
    L_i=\min_l\ \{l\ |\ H^{(l)}_{{\rm Rx},i}<\eta\}.
\end{align} 
Accordingly, the adaptive detection rule is given by
\begin{align}\label{eq: adaptive iteration2}
    \hat{w}_i^{(l)}= 
    \begin{cases}
        \hat{w}_i^{(l-1)},&{\rm if}\ i\notin \mathcal{A}^{(l)},\\
        \underset{w_i}{\rm{argmax}}\ P({\bf y}_i|w_i)\hat{P}(w_i|\hat{\mathbf{w}}^{(l-1)}_{\backslash i}),&{\rm if}\ i\in \mathcal{M}^{\rm c}\cap\mathcal{A}^{(l)},\\
        \underset{w_i}{\rm{argmax}}\ \hat{P}(w_i|\hat{\mathbf{w}}^{(l-1)}_{\backslash i}),&{\rm if}\ i\in \mathcal{M}\cap\mathcal{A}^{(l)},
    \end{cases}
\end{align}
where the active token set is formulated as
\begin{align}\label{eq: adaptive iteration3}
    \mathcal{A}^{(l)}=
    \begin{cases}
    \{1,...,T \},&\ \text{if }l=0,\\
    \{i\ |\ H^{(l-1)}_{{\rm Rx},i}\geq\eta\},&\ \text{if }l>0.
    \end{cases}
\end{align}

As the detection proceeds, $|\mathcal{A}^{(l)}|$ monotonically decreases since more tokens reach sufficient confidence and become frozen.
Further, tokens that are easily inferred, either because the channel likelihood is highly reliable or because their contextual prior is already sharply peaked, reach the entropy threshold early and terminate their updates. 
Consequently, computational resources are concentrated on uncertain tokens, reducing the number of unnecessary MLM evaluations.

\vspace{6mm}
\section{Tx Strategy of the Proposed Framework: Context-Aware Token Masking}

Building upon the Rx-side context-aware detection strategy, this section develops a context-aware token masking strategy at the Tx.
Instead of transmitting all tokens equally important, the Tx selectively omits tokens that can be reliably inferred at the Rx even without the channel observation.
In principle, the Tx-side masking strategy aims to select a masking set $\mathcal{M}\subset\{1,...,T\}$ that maximizes the accuracy of token reconstruction under the total power constraint $P_{\rm tot}$.
To derive a tractable solution, we decompose the Tx strategy into two coupled components: 1) sequential masking selection based on the prior entropy (Sec. V-A), and 2) masking ratio control based on token detection probability comparison (Sec. V-B).

\subsection{Sequential Masking Selection based on Prior Entropy}\label{subSec: Greedy Entropy}
According to \eqref{eq: modified iteration}, the token detection rule at iteration $l$ for masked tokens is given by
\begin{align}\label{eq: modified iteration2}
    \hat{w}_i^{(l)}=
     \underset{w_i}{\rm{argmax}}\ \hat{P}(w_i|\hat{\mathbf{w}}^{(l-1)}_{\backslash i}),\ \forall i\in \mathcal{M}.
\end{align}
This indicates that the detection of masked tokens relies solely on the contextual prior derived from previously detected tokens. Accordingly, our objective is to quantify the \emph{confidence level} of each token estimate given the surrounding tokens.

To this end, similar to the token-wise iteration control in Sec.~IV-B, we consider the prior entropy $H^{(l)}_{{\rm Rx},i}$ defined in \eqref{eq: adaptive iteration entropy},
which quantifies the uncertainty of the  model’s prior output and the resulting detection uncertainty for the masked tokens.
However, computing $H^{(l)}_{{\rm Rx},i}$ requires knowledge of $\hat{\mathbf{w}}^{(l-1)}_{\backslash i}$, which is not available at the Tx.
Without this information, evaluating $\hat{P}(w_i|\hat{\mathbf{w}}^{(l-1)}_{\backslash i})$ at the Tx would require marginalization over all possible realizations of $\hat{\mathbf{w}}^{(l-1)}_{\backslash i}$, involving $V^{T-1}$ combinations.

To address this issue, we approximate $\hat{\mathbf{w}}^{(l-1)}$ by the true masked token sequence $\mathbf{w}_{\rm m} = {\rm Mask}(\mathbf{w};\mathcal{M})$ when optimizing the masking selection at the Tx. Notably, if all the unmasked tokens are correctly detected at the Rx, then $\hat{\mathbf{w}}^{(l-1)}$  and $\mathbf{w}_{\rm m}$ coincide at all unmasked positions. Hence, $\mathbf{w}_{\rm m}$ can be interpreted as an idealized version of $\hat{\mathbf{w}}^{(l-1)}$ for these positions. Based on this approximation, we define the prior entropy at the Tx as
\begin{align}
    H_{{\rm Tx},i}=\!-\!\underset{v}{\sum}\hat{P}(w_i\!=\!v|\mathbf{w}_{{\rm m},\backslash i})\text{log}_2\hat{P}(w_i\!=\!v|\mathbf{w}_{{\rm m},\backslash i}),\forall i\in\mathcal{M},
\end{align}
where $\mathbf{w}_{{\rm m},\backslash i}$ denotes the masked token sequence excluding the $i$-th token.
Consequently, we formulate the problem of mask selection that minimizes the prior entropy $H_{{\rm Tx},i}$ as follows:
\begin{align}
    \underset{\mathcal{M}}{\rm{argmin}}&\ \sum_{i\in\mathcal{M}} H_{{\rm Tx},i} \nonumber \\
    \text{s.t.}~~~&|\mathcal{M}| = N.
\end{align}

Directly solving this problem is computationally challenging, as it requires a joint search over all possible masking patterns, leading to a combinatorial optimization problem. To reduce the complexity, we adopt a sequential greedy masking strategy based on contextual priors at the Tx. At the $n$-th masking step $(n \ge 0)$, given the masked sequence $\mathbf{w}_{\rm m}^{(n)}$ and masking set $\mathcal{M}^{(n)}$, the Tx evaluates the contextual prior distribution of each unmasked token using the MLM:
\begin{align}\label{eq: MLM output}
\hat{P}(w_i|\mathbf{w}^{(n)}_{{\rm m},\backslash i})=[{\rm MLM}(\mathbf{w}^{(n)}_{{\rm m},\backslash i})]_i.
\end{align}
Based on this prior, the prior entropy measure at the Tx is defined below.
\vspace{2mm}\\
{\bf Definition 2} (Prior Entropy of Token Masking).
{\em The contextual uncertainty of the $i$-th token at the $n$-th masking step is defined as the entropy of the MLM output
\begin{align}\label{eq: entropy for masking}
H_{{\rm Tx},i}^{(n)}\triangleq-\sum_v \hat{P}(w_i=v|\mathbf{w}^{(n)}_{{\rm m},\backslash i}){\rm log}_2\hat{P}(w_i=v|\mathbf{w}^{(n)}_{{\rm m},\backslash i}).
\end{align}
which measures the uncertainty of the detection for masking of that token given the current masked token sequence.}\\
Using this as the masking score, the next token to be masked is selected as the most predictable one, i.e., the token with the smallest entropy:
\begin{align}\label{eq: greedy selection}
i^*_{n} = \underset{i\notin \mathcal{M}^{(n)}}{\text{argmin}}\ {H}_{{\rm Tx},i}^{(n)}.
\end{align}
Accordingly, the masking set and masked sequence are updated as $\mathcal{M}^{(n+1)}=\mathcal{M}^{(n)}\cup\{i^*_n\}$ and $\mathbf{w}_{\rm m}^{(n+1)}={\rm Mask}(\mathbf{w};\mathcal{M}^{(n+1)})$. By construction, $\mathbf{w}^{(0)}_{\rm m}=\mathbf{w}$ and $\mathcal{M}^{(0)}=\varnothing$. A lower entropy indicates that the token can be inferred from context with high confidence and is thus a strong candidate for masking, whereas high-entropy tokens are prioritized for transmission.

\subsection{Masking Ratio Control}\label{subSec: r determine}
In this subsection, we determine the masking ratio, defined as $r = N/T$, where $N = |\mathcal{M}|$ is the number of masked tokens. This is equivalent to determining the stopping point of the sequential masking selection process described in Sec. V-A.
To this end, we begin by analyzing the fundamental trade-off underlying the detection rule in \eqref{eq: modified iteration}. As shown in \eqref{eq: received2}, masking $N$ tokens leads to allocating the total transmit power $P_{\rm tot}$ over the remaining $T-N$ transmitted tokens. As a result, the detection reliability of unmasked tokens improves with increasing $N$, due to the higher effective SNR. However, increasing $N$ also reduces the number of observed (unmasked) tokens available as context, which can degrade the quality of the contextual prior for masked tokens at the Rx. Therefore, the masking ratio $r$ should be chosen to balance this trade-off between observation reliability and contextual prior.

To formally characterize this trade-off, we consider two scenarios at the $n$-th step of the sequential masking procedure (Sec.~V-A), where the Tx decides whether to mask $w_{i^*_n}$ or to transmit it.
In the first scenario, the token $w_{i^*_n}$ is masked, while the remaining tokens $w_{i^*_{n+1}},\ldots,w_{i^*_{T-1}}$ are transmitted without masking. Consequently, the total number of masked tokens is given by $|\mathcal{M}| = n + 1$. In this scenario, the probability of correctly detecting tokens $w_{i^*_{n}},\ldots,w_{i^*_{T-1}}$ is given by
\begin{align} 
    &P_{{\rm d},0}(n)\!\nonumber\\&\!\!\triangleq P\Big((\hat{w}_{i^*_n},...,\hat{w}_{i^*_{T-1}})\!=\!(w_{i^*_n},...,w_{i^*_{T-1}}); |\mathcal{M}|\!=\!n\!+\!1\Big) \nonumber\\
    &\!\!= P\Big((\hat{w}_{i^*_{n+1}},...,\hat{w}_{i^*_{T-1}})\!=\!(w_{i^*_{n+1}},...,w_{i^*_{T-1}}); |\mathcal{M}|\!=\!n\!+\!1\Big) \nonumber \\
    &~\times P\Big(\hat{w}_{i^*_{n}}\!=\!w_{i^*_{n}}~\big|~ (\hat{w}_{i^*_{n+1}},...,\hat{w}_{i^*_{T-1}})\!=\!(w_{i^*_{n+1}},...,w_{i^*_{T-1}})\nonumber\\&~~~~~~~~~~~~~~~~~~~~~~~~~~~~~~~~~~~~~~~~~~~~~~~~~~~~~~;|\mathcal{M}|\!=\!n\!+\!1\Big) \nonumber \\
    &\!\!\approx P\Big((\hat{w}_{i^*_{n+1}},...,\hat{w}_{i^*_{T-1}})\!=\!(w_{i^*_{n+1}},...,w_{i^*_{T-1}}); |\mathcal{M}|\!=\!n\!+\!1\Big) \nonumber \\
    &~\times \hat{P}(w_{i^*_n}| \mathbf{w}^{(n)}_{{\rm m},\backslash i^*_n}),
\end{align}
where $\hat{w}_i$ denotes the detected token at the Rx. Also, the last approximation assumes that the probability of correctly inferring the masked token $w_{i^*_n}$ follows the MLM-derived prior.
In the second scenario, all unmasked tokens $w_{i^*_{n}},\ldots,w_{i^*_{T-1}}$ are transmitted without masking. Consequently, the total number of masked tokens is given by $|\mathcal{M}|=n$. In this scenario, the probability of correctly detecting $w_{i^*_{n}},\ldots,w_{i^*_{T-1}}$ is given by
\begin{align}
    P_{{\rm d},1}(n)\!\triangleq\! P\Big((\hat{w}_{i^*_n},...,\hat{w}_{i^*_{T-1}})\!=\!(w_{i^*_n},...,w_{i^*_{T-1}}); |\mathcal{M}|\!=\!{n}\Big).
\end{align}

To further characterize the probabilities $P_{{\rm d},0}(n)$ and $P_{{\rm d},1}(n)$, we employ an ML-based BER expression as a tractable surrogate for the MAP-based BER. This is because an exact BER characterization under MAP detection would require marginalization over all possible token-sequence combinations of $w_{i^*_n},...,w_{i^*_{T-1}}$, which is computationally intractable.
Under the assumption that the MLM provides sufficiently reliable priors, the MAP detector is expected to outperform its ML counterpart. Accordingly, the ML-based BER provides a tractable lower-bound approximation of $P_{\rm d,0}(n)$ and $P_{\rm d,1}(n)$. The resulting expressions are given as follows:
\begin{align}
    P_{\rm d, 0}(n)&\geq\left(1-P_b\left(\frac{P_{\rm tot}}{T\!-\!n\!-\!1}\right)\right)^{(T-n-1)\lceil{\rm log}_2(V)\rceil}\nonumber\\&\quad \quad \quad \quad \quad \quad \quad~\times \hat{P}(w_{i^*_n}| \mathbf{w}^{(n)}_{{\rm m},\backslash i^*_n}) \triangleq \hat{P}_{{\rm d},0}(n),\label{eq: P2} \\
    P_{\rm d, 1}(n)&\geq\left(1-P_{b}\left(\frac{P_{\rm tot}}{T-n}\right)\right)^{(T-n)\lceil{\rm log}_2(V)\rceil}\!\!\!\!\!\triangleq \hat{P}_{{\rm d},1}(n),\label{eq: P1}
\end{align}
where $P_b(\cdot)$ is the BER function of ML detection for $2^m$-QAM modulation, which is approximated as \cite{BER_approx}:
\begin{align}\label{eq: BER}
    P_b(p)\approx &\frac{\sqrt{2^m}-1}{\sqrt{2^m}\log_2\sqrt{2^m}}{\rm erfc}\left(\sqrt{\frac{3p\gamma}{2(2^m-1)}}\right)\nonumber\\
    &+\frac{\sqrt{2^m}-2}{\sqrt{2^m}\log_2\sqrt{2^m}}{\rm erfc}\left(3\sqrt{\frac{3p\gamma}{2(2^m-1)}}\right),
\end{align}
where $p$ denotes the symbol power, $\gamma=\frac{|h|^2}{\sigma^2}$ represents the channel-gain-to-noise-power ratio, and $\text{erfc}(x)=1-\frac{2}{\sqrt{\pi}}\int_{0}^x\text{exp}(-u^2){\rm d}u$ denotes the complementary error function.
In this work, we assume perfect channel state information (CSI) at the Tx, such that $\gamma$ can be directly computed.

\begin{figure}[t]
    {\epsfig{file=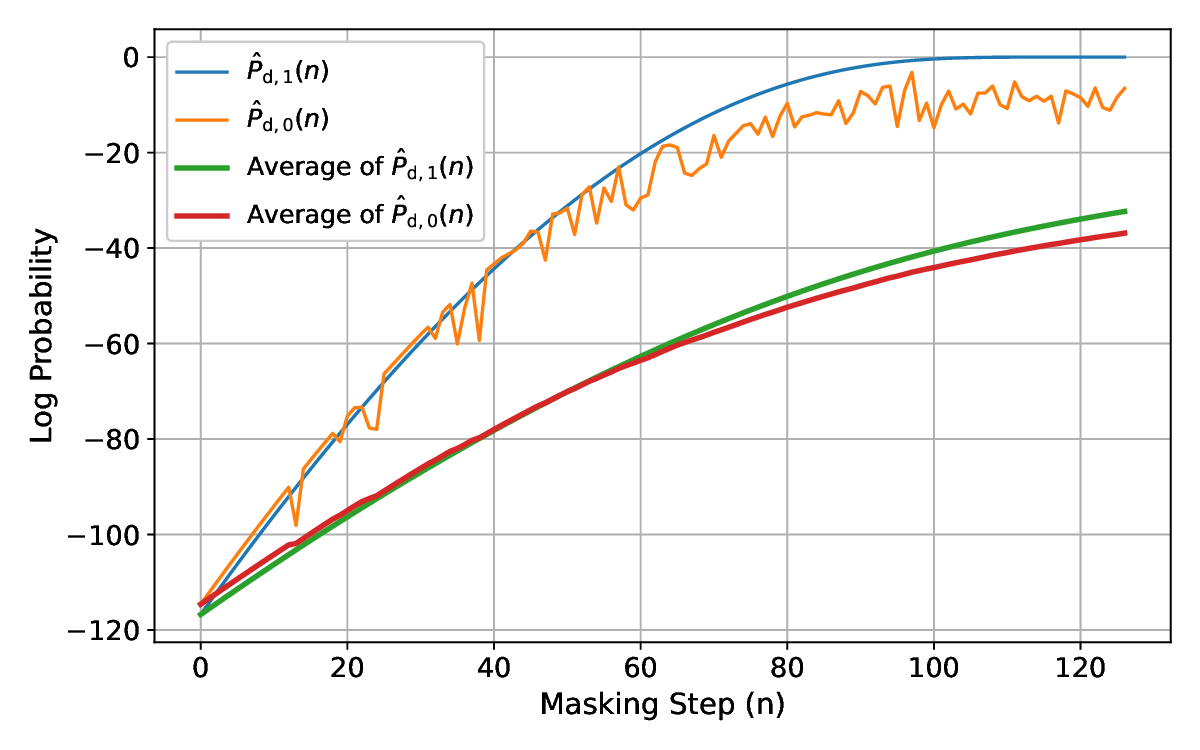,width=8.5cm}} 
    \vspace{-3mm}
    \caption{An illustration of instantaneous and averaged log probabilities of $\hat{P}_{{\rm d},1}(n)$ and $\hat{P}_{{\rm d},0}(n)$ ($T=128$, 16-QAM, instantaneous SNR $=10$ dB). The harmonic mean suppresses local fluctuations in $\hat{P}_{{\rm d},0}(n)$ caused by sequential context-prior variations, revealing a well-defined stopping point for masking.}
    \vspace{-3mm} 
    \label{Fig: log prabability}
\end{figure}

 These analytical results imply that when $\hat{P}_{{\rm d},1}(n)<\hat{P}_{{\rm d},0}(n)$, masking $w_{i^*_n}$ yields a higher detection probability than transmitting it, and thus the masking process should continue. Conversely, when $\hat{P}_{{\rm d},1}(n)>\hat{P}_{{\rm d},0}(n)$, further masking becomes detrimental, and the masking process should be terminated.
Based on this observation, we investigate the existence of a stopping point for the masking process. To this end, we adopt the assumption stated in \textbf{Assumption 1}, which characterizes the degradation of the contextual prior as masking progresses.
\vspace{2mm} \\
{\bf Assumption 1} (Monotonic Degradation of Contextual Prior under Sequential Masking). {\em As masking progresses and less context remains available, the reliability of inferring a newly masked token is assumed to be non-increasing with the masking index. Formally,}
\begin{align}
    \hat{P}(w_{i^*_n}| \mathbf{w}^{(n)}_{{\rm m},\backslash i^*_n})\leq \hat{P}(w_{i^*_{n-1}}| \mathbf{w}^{(n-1)}_{{\rm m},\backslash i^*_{n-1}}).
\end{align}

Under this assumption, the prior term in $\hat{P}_{{\rm d},0}(n)$ decreases with $n$.
Consequently, a stopping point beyond which additional masking is no longer beneficial is expected to emerge.
Based on {\bf Assumption 1}, masking decisions are made by comparing token detection probabilities $\hat{P}_{{\rm d},1}$ and $\hat{P}_{{\rm d},0}$. This comparison leads to two operational criteria, namely the {\em 1) Instantaneous probability criterion} and the {\em 2) Average probability criterion} described next.

\subsubsection{Instantaneous probability criterion}
Masking is continued while contextual inference remains more reliable than transmission, leading to the stopping index
\begin{align}\label{eq: criterion1}
    N=\min_n\ \{n\ |\ \hat{P}_{\rm d,1}(n)>\hat{P}_{\rm d,0}(n)\}.
\end{align}
This rule identifies the transition point where transmitting the selected token becomes more reliable than masking it.
However, the monotonic behavior implied by {\bf Assumption 1} may not strictly hold in practice.
As observed in Fig. \ref{Fig: log prabability}, the detection probability $\hat{P}_{\rm d,0}(n)$ may fluctuate across masking steps due to the entropy-based sequential selection process. Therefore, direct comparison between $\hat{P}_{\rm d,1}(n)$ and $\hat{P}_{\rm d,2}(n)$ may therefore become unstable.
To obtain a more stable operational criterion, we next consider a probability measure that captures the cumulative masking trend rather than instantaneous variations.

\subsubsection{Average probability criterion}
In this approach, we aggregate the instantaneous probabilities using the geometric mean:
\begin{align}\label{eq: p_hat}
    \bar{P}_{{\rm d},a}(n)\triangleq \left(\prod_{i=1}^n\hat{P}_{{\rm d},a}(i)\right)^{\frac{1}{n}},\ \forall a\in\{0,1\}. 
\end{align}
The geometric mean is adopted instead of the arithmetic mean due to the exponential dependence on $n$ in the definitions of $\hat{P}_{\rm d,1}(n)$ and $\hat{P}_{\rm d,0}(n)$. Using an arithmetic mean would cause the term corresponding to $i = n$ to dominate the sum, thereby failing to capture the overall trend of the cumulative probability. In contrast, the geometric mean computes an average in the log-domain, providing a more balanced measure.
As shown in Fig. \ref{Fig: log prabability}, the curves of $\bar{P}_{\rm d,1}(n)$ and $\bar{P}_{\rm d,0}(n)$ evolve smoothly and exhibit a clear intersection point. Accordingly, the masking process is terminated at the smallest index $N$ satisfying
\begin{align}\label{eq: criterion2}
    N=\min_n\left\{n\ |\ \bar{P}_{{\rm d},1}(n)>\bar{P}_{{\rm d},0}(n) \right\}.
\end{align}
This stopping rule determines the masking ratio based on cumulative probability rather than instantaneous comparison, yielding a robust and stable operational decision under dynamically updated contextual priors.

\begin{table}[t]
\centering
\caption{SIM Performance and Corresponding Average Masking Ratios of the Proposed Framework with Different Masking Criterion Using the Europarl corpus dataset \cite{Europarl}.}\label{table: criterion}
\resizebox{\columnwidth}{!}{%
{\small
\begin{tabular}{ccccc}
\hline
\multicolumn{5}{c}{\text{\begin{tabular}[c]{@{}c@{}}SIM\\ (Average masking ratio $r$)\end{tabular}}} \\ \hline
\multicolumn{1}{c||}{\text{Criterion}} & \multicolumn{1}{c|}{\text{0 dB}} & \multicolumn{1}{c|}{\text{5 dB}} & \multicolumn{1}{c|}{\text{10 dB}} & \text{15 dB} \\ \hline
\multicolumn{1}{c||}{1) Instantaneous \eqref{eq: criterion1}} & \multicolumn{1}{c|}{\begin{tabular}[c]{@{}c@{}}0.3460\\ (0.1991)\end{tabular}} & \multicolumn{1}{c|}{\begin{tabular}[c]{@{}c@{}}0.6193\\ (0.1707)\end{tabular}} & \multicolumn{1}{c|}{\begin{tabular}[c]{@{}c@{}}0.8018\\ (0.1245)\end{tabular}} & \begin{tabular}[c]{@{}c@{}}0.9183\\ (0.0609)\end{tabular} \\ \hline
\multicolumn{1}{c||}{\bf 2) Average \eqref{eq: criterion2}} & \multicolumn{1}{c|}{\textbf{\begin{tabular}[c]{@{}c@{}}0.4198\\ (0.7790)\end{tabular}}} & \multicolumn{1}{c|}{\textbf{\begin{tabular}[c]{@{}c@{}}0.6987\\ (0.5133)\end{tabular}}} & \multicolumn{1}{c|}{\textbf{\begin{tabular}[c]{@{}c@{}}0.8554\\ (0.2850)\end{tabular}}} & \textbf{\begin{tabular}[c]{@{}c@{}}0.9423\\ (0.1223)\end{tabular}} \\ \hline
\end{tabular}%
}
}
\vspace{-3mm}
\end{table}

{\small \RestyleAlgo{ruled}
    \SetKwComment{Comment}{/* }{ */}
    \begin{algorithm}[t]\label{alg:Tx Strategy}
    \caption{Procedure of the proposed framework}
    \textbf{1. Context-Aware Token Masking}:\\
    ${\bf w}_{\rm m}^{(0)}={\bf w}$;\\
    $\mathcal{M}^{(0)}=\varnothing$;\\
        \For{$n\in\{0,...,T-1\}$}{
            Get $P(w_i|\mathbf{w}^{(n)}_{{\rm m},\backslash i}),\ \forall i\notin\mathcal{M}^{(n)}$ in \eqref{eq: MLM output};\\
            Calculate $H_{{\rm Tx},i}^{(n)},\ \forall i\notin\mathcal{M}^{(n)}$ in \eqref{eq: entropy for masking};\\
            
            $i^*_{n} = \underset{i\notin \mathcal{M}^{(n)}}{\text{argmin}}\ H_{{\rm Tx},i}^{(n)}$;\\

            Calculate $\bar{P}_{\rm d,1}(n),\bar{P}_{\rm d,2}(n)$ in \eqref{eq: P2}, \eqref{eq: P1}, \eqref{eq: p_hat};\\
            \textbf{if} $\bar{P}_{\rm d,1}(n)>\bar{P}_{\rm d,2}(n)$ \textbf{then break};\\
            
            $\mathcal{M}^{(n+1)}=\mathcal{M}^{(n)}\cup\{i^*_n\}$;\\
            $\mathbf{w}_{\rm m}^{(n+1)}={\rm Mask}(\mathbf{w};\mathcal{M}^{(n+1)})$;\\
        }
        $\mathcal{M}=\mathcal{M}^{(n)}$;\\
        $N=n$;
        \vspace{2mm}\\
        \textbf{2. Wireless Transmission}:\\
         Map $w_i$ into complex symbol ${\sf s}_i(w_i),\ \forall i\notin\mathcal{M}$
        ${\bf y}_i=h\sqrt\frac{P_{\rm tot}}{T-N}{\sf s}_i(w_i)+{\bf n}_i,\ \forall i\notin \mathcal{M}$;
        \vspace{2mm}\\
        \textbf{3. Context-Aware Token Detection}:\\
        $P(w_i|\mathbf{\hat w}_{\backslash i}^{(-1)})=\frac{1}{V},\ \forall i$;\\
    Calculate $P({\bf y}_i|w_i),\ \forall i\notin\mathcal{M}$ as in \eqref{eq: modified2};\\
    $\mathcal{A}^{(0)}=\{1,...,T\}$;\\
    \For{$l\in\{0,...,L_{\rm max}\}$}{
        $\hat{w}_i^{(l)}\!\!=\! \underset{w_i}{\text{argmax}}~P({\bf y}_i|w_i)P(w_i|\mathbf{\hat w}_{\backslash i}^{(l-1)}),\forall i\!\in\! \mathcal{M}\cap\mathcal{A}^{(l)}$;\\
        $\hat{w}_i^{(l)}\!\!=\! \underset{w_i}{\text{argmax}}~P(w_i|\mathbf{\hat w}_{\backslash i}^{(l-1)}),\forall i\!\in\! \mathcal{M}^{\rm c}\!\cap \mathcal{A}^{(l)}$;\\
        $\hat{w}_i^{(l)}= \hat{w}_i^{(l-1)},\ \forall i\notin\mathcal{A}^{(l)}$;\\
        Calculate $H^{(l)}_{{\rm Rx},i},\ \forall i$ as in \eqref{eq: adaptive iteration entropy};\\
        $\mathcal{A}^{(l+1)}=\{i\ |\ H^{(l)}_{{\rm Rx},i}\geq\eta\}$;\\
        \textbf{if} $|\mathcal{A}^{(l+1)}|=0$ \textbf{then break};\\
        \textbf{if} $l=0$ \textbf{then } $\hat{w}_{i^*}^{(0)} = [{\rm MASK}], \forall i^*\in\mathcal{M}$;\\
        $P(w_i|\mathbf{\hat w}_{\backslash i}^{(l)})=[{\rm MLM}(\mathbf{\hat w}_{\backslash i}^{(l)})]_i,\ \forall i\in\mathcal{A}^{(l+1)}$;\\
        
    }
    \end{algorithm}
    }

To compare the instantaneous probability criterion in \eqref{eq: criterion1} and the average probability criterion in \eqref{eq: criterion2}, we evaluate the contextual fidelity of the reconstructed text using {\em SIM} performance, defined as the cosine similarity between text embeddings of the original and reconstructed texts. Detailed description of the SIM metric is provided in Sec. VI. Table \ref{table: criterion} reports the resulting SIM values together with the corresponding average masking ratios $(r)$ across SNR levels for 4-QAM modulation.
Compared to the instantaneous probability criterion, the average probability criterion consistently achieves higher SIM performance, demonstrating more stable and reliable masking decisions.
This is because fluctuations in $\hat{P}_{\rm d,0}(n)$ can cause \eqref{eq: criterion1} to be satisfied at early masking steps, preventing the selection of an appropriate masking ratio and resulting in smaller values of $r$.
Based on these results, we adopt the average probability criterion in \eqref{eq: criterion2} for masking ratio control in the sequential masking. The procedure of the proposed framework introduced in Sec. IV, V is summarized in Algorithm \ref{alg:Tx Strategy}.

\section{Simulation Results}
\begin{figure*}[t]
    \centering
    \subfigure[The Europarl corpus dataset.]{\epsfig{file=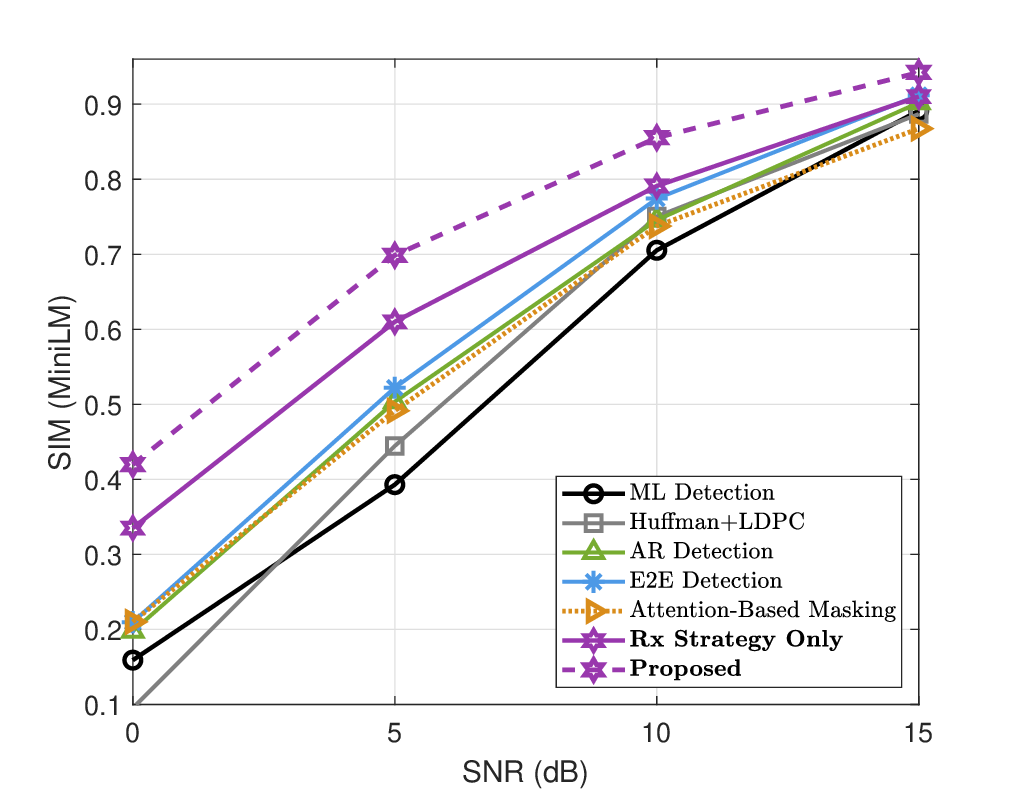,width=8cm} 
    \epsfig{file=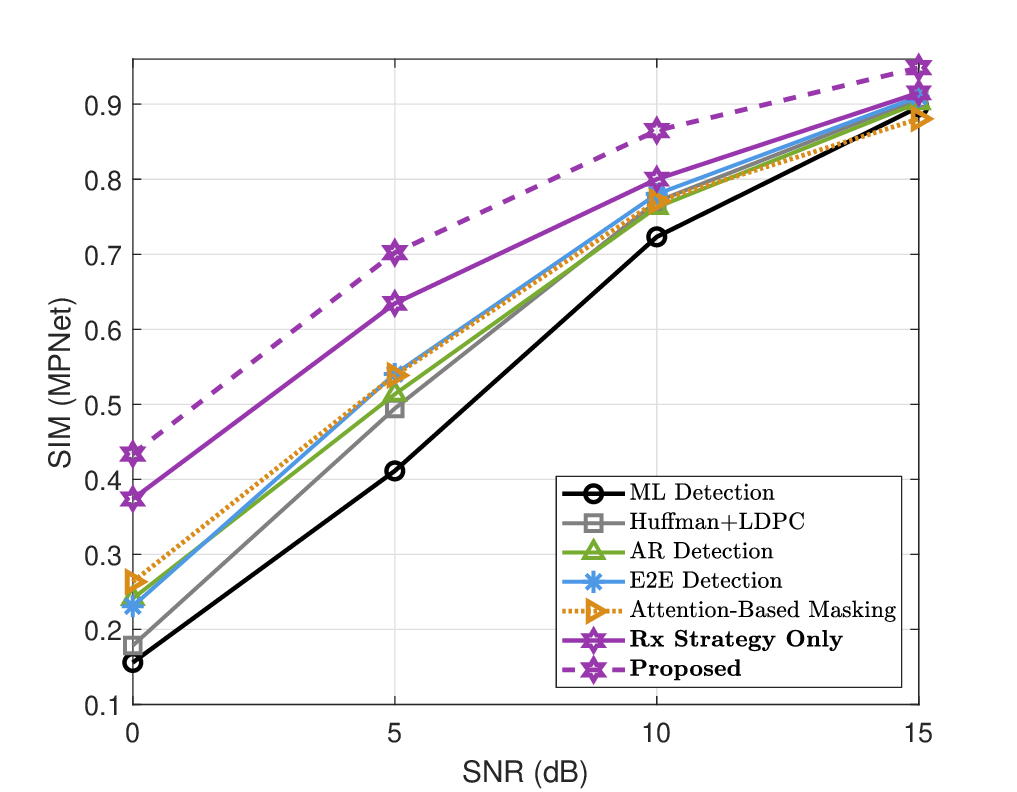,width=8cm}}
    \subfigure[The WikiText-103 dataset.]
    {\epsfig{file=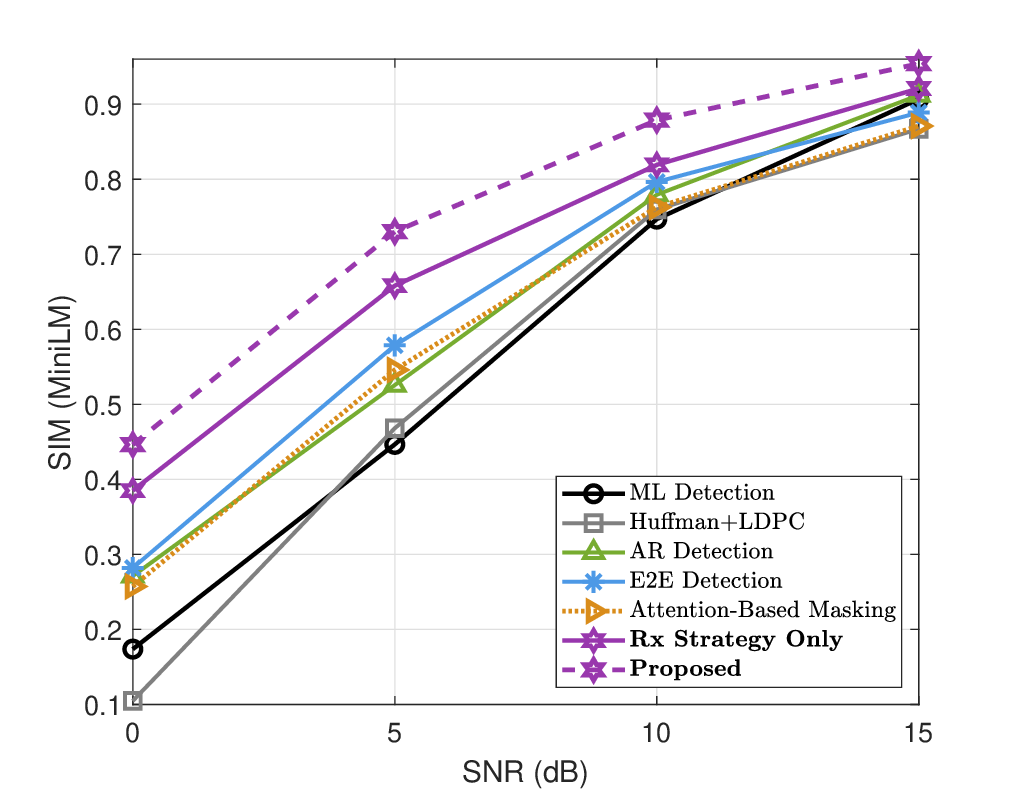,width=8cm}
    \epsfig{file=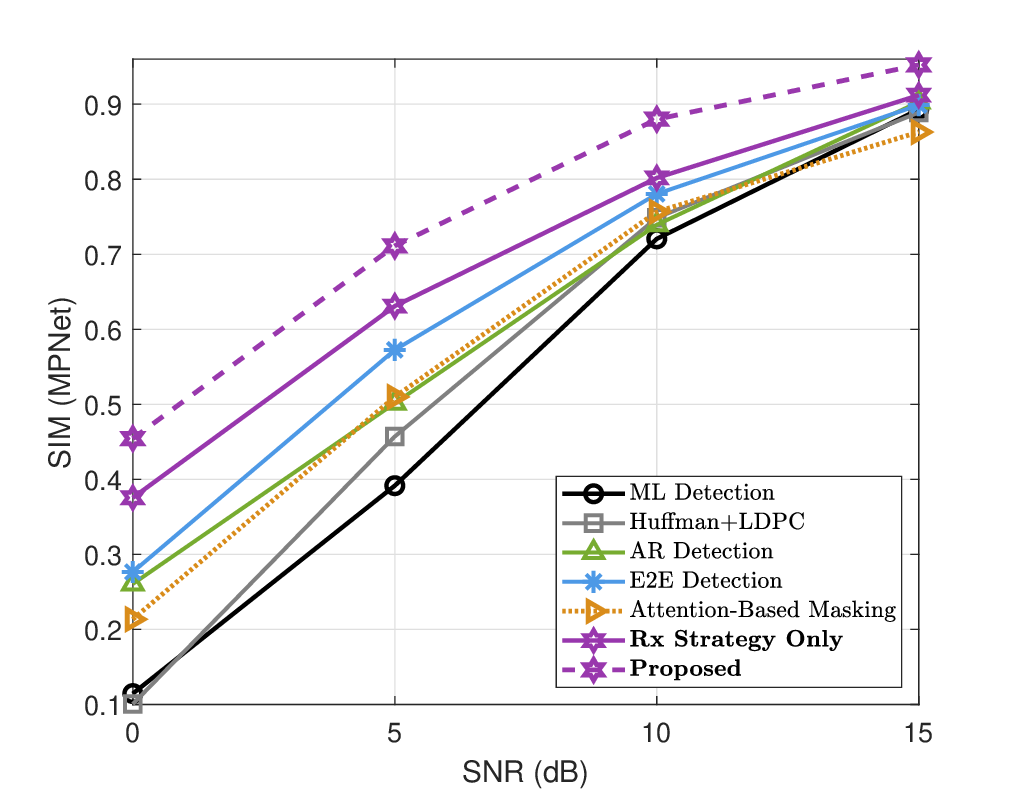,width=8cm}}
    \caption{SIM performance of the various token communication frameworks across different SNRs.}\vspace{-3mm} 
    \label{Fig: Overall}
\end{figure*}
We evaluate the proposed context-aware token communication framework using a text transmission task.
The Europarl corpus \cite{Europarl} and the WikiText-103 dataset \cite{WikiText-103} are used as text datasets, and tokens are generated by WordPiece-based tokenization \cite{WordPiece}. The BERT model \cite{BERT} serves as the shared MLM at both the Tx and Rx.
We consider packets consisting of  $T=128$ tokens, each represented by 15 bits. For the physical-layer modulation, 4-QAM is applied over a Rayleigh block fading channel.
To assess the contextual fidelity of the reconstructed text, we compute the cosine similarity between text embeddings of the original and reconstructed texts, as considered in \cite{TokComm_packet,TokComm_crossmodal,evaluation_metric,evaluation_metric2}. Specifically, the two texts, each consisting of $T$ tokens, are encoded using a sentence-transformer model to obtain their embedding vectors. The cosine similarity between the resulting embeddings is then used as the evaluation metric, denoted as {\em SIM}. In this work, we use the MiniLM \cite{sentence_embedding} and MPNet \cite{sentence_embedding_mpnet} sentence-transformer model to generate the embeddings.
For the proposed framework, the maximum number of prior update is set to $L_{\rm max}=5$ and the posterior-entropy threshold used for token-wise iteration control is set to $\eta = 2\times 10^{-3}$.

\subsection{Reconstruction Performance of the Proposed Framework}

In this subsection, we evaluate the effectiveness of the proposed context-aware token communication framework. For performance comparison, we consider the following token communication frameworks:
\begin{itemize}
    \item {\bf ML Detection:} This is a conventional physical-layer baseline that estimates tokens solely from the channel likelihood, without incorporating any contextual prior across tokens. Concretely, each token is detected independently by maximizing the channel likelihood term as \eqref{eq: ML detection}.
    
    \item {\bf Huffman+LDPC\cite{Huffman,LDPC}:} This is a conventional separated source–channel coding baseline. A Huffman codebook \cite{Huffman} is constructed from the WikiText-103 training data to encode tokens into a bitstream, and the resulting bitstream is protected by a 4/5-rate low-density parity check (LDPC) code \cite{LDPC}. If the total overhead exceeds the budget $T\lceil{\rm log}_2(V)\rceil$ bits, tokens generating the longest codewords are progressively dropped to meet the overhead constraint.

    \item {\bf AR Detection\cite{LLM-SC,GPT2}:} This is an autoregressive (AR) token detection baseline suggested in \cite{LLM-SC}, using a GPT-2–based \cite{GPT2} AR language model trained on the WikiText-103 training data. Detection is formulated in an AR fashion using the prior derived from the model as follows:
    \begin{align}
        \hat{w}_i=\underset{w_i}{\text{argmin}}\ P(\mathbf{y}_i|w_i)P(w_i|\hat{w}_1,...,\hat{w}_{i-1}).
    \end{align}
    In terms of model complexity, the AR detection framework employs 132.0 million parameters, whereas the proposed framework requires 109.5 million parameters.

    \item {\bf E2E Detection\cite{Semantic_Source_Channel}:} This is an end-to-end (E2E) learned detection baseline \cite{Semantic_Source_Channel}, where a neural model is trained to map the initial ML token estimates to a refined token sequence in a single shot. The same model structure with the proposed framework is deployed, and the model is trained using the WikiText-103 training data, aiming to learn a direct correction mapping without any explicit Bayesian formulation.

    \item {\bf Attention-Based Masking\cite{attention, attention_Jiwoong}:} This is a masking-based baseline suggested in prior literature. Tokens are selected for masking according to their attention scores \cite{attention} obtained from the MLM, following the approach suggested in \cite{attention_Jiwoong}. Tokens with lower attention scores are selected for masking. The masked tokens are then inferred at the Rx based on the ML detection results of the unmasked tokens. The masking ratio is fixed to $r=0.3$.

    \item {\bf Rx Strategy Only:} This is a detection baseline that applies the proposed context-aware token detection at the Rx without employing the Tx masking strategy. In this configuration, all tokens are transmitted and the Rx performs iterative MAP-based token detection using a shared MLM, as described in Sec. IV. For this configuration, the posterior-entropy threshold for token-wise iteration control is set to $\eta=5\times 10^{-4}$.

\end{itemize}

Fig. \ref{Fig: Overall} presents the SIM performance of different token communication frameworks. The results show that the proposed token communication framework consistently achieves the highest SIM performance compared to all considered baseline schemes across the entire SNR range. In particular, the proposed framework attains higher SIM performance than conventional physical-layer-driven schemes (e.g., ML Detection and Huffman+LDPC) as well as learning-based detection schemes (e.g., AR Detection and E2E Detection).
It also achieves higher SIM performance than the Attention-Based Masking baseline that relies on attention scores for token masking and reconstruction.
Furthermore, incorporating the proposed Tx-side masking strategy into the proposed framework yields additional SIM performance gains, where the joint Tx-Rx configuration further improves the reconstruction accuracy compared to the Rx Strategy Only configuration.
These results demonstrate the effectiveness of the proposed context-aware token communication framework in exploiting contextual dependencies among tokens for robust context recovery.

\subsection{Impact of the Tx Masking}
In this subsection, we evaluate the impact of different Tx-side masking strategies on the reconstruction performance. The same Rx strategy is applied in all experiments in this subsection, and the maximum number of MLM refinements is fixed to $L_{\rm max}=5$. For comparison, we consider the following Tx masking strategies.
\begin{itemize}
    \item {\bf Random Masking (fixed $r$):} Tokens are masked uniformly at random using a fixed masking ratio $r$, without accounting for their predictability at the Rx. This serves as a baseline that ignores context structure at the Tx.
    \item {\bf Parallel Masking (fixed $r$):} This strategy is a non-iterative variant of the proposed sequential masking policy. Instead of sequentially updating the masking decisions, the Tx selects the $T\times r$ tokens with the lowest $H^{(0)}_{{\rm Tx},i}$ in a single step, with the masking ratio $r$ fixed in advance. As a result, the masking process cannot incorporate updated contextual information revealed after each masking step.  
    \item {\bf Sequential Masking (fixed $r$):} Tokens are selected for masking using the proposed entropy-based sequential policy developed in Sec. V-A, while the masking ratio $r$ is fixed in advance. This strategy exploits contextual predictability but still lacks adaptability to varying channel conditions and token instances.
\end{itemize}
\begin{figure}[t]
    \centering
    \subfigure[The Europarl corpus dataset.]
    {\epsfig{file=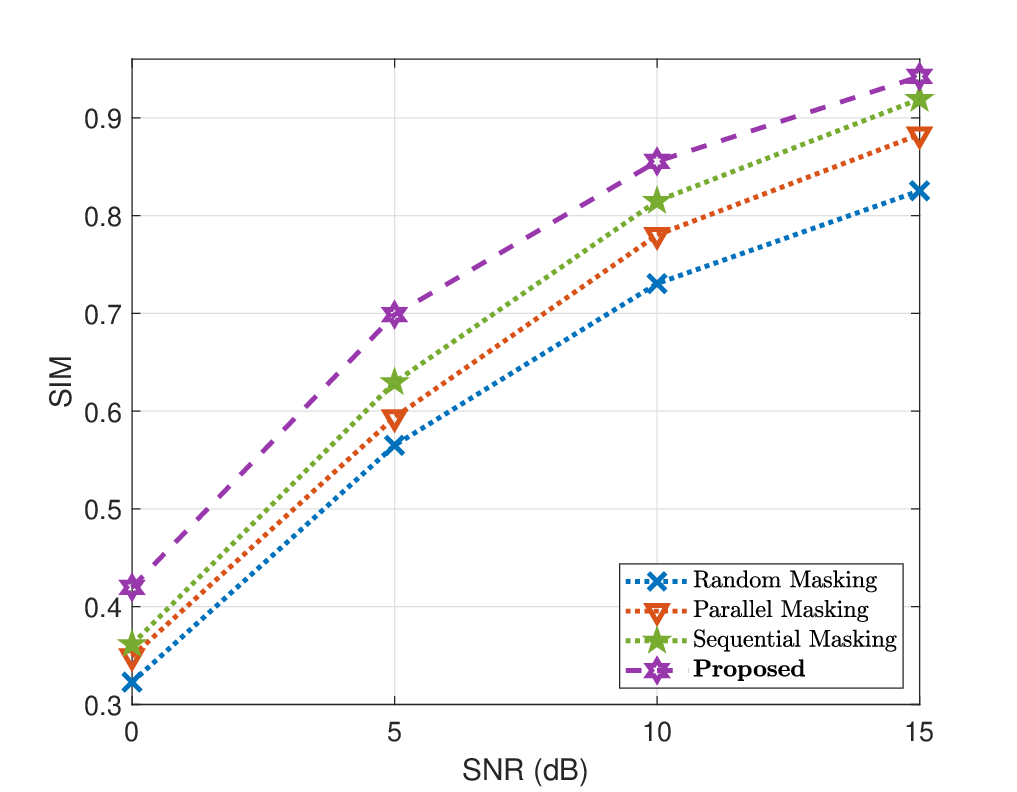,width=8cm}}
    \subfigure[The WikiText-103 dataset.]
    {\epsfig{file=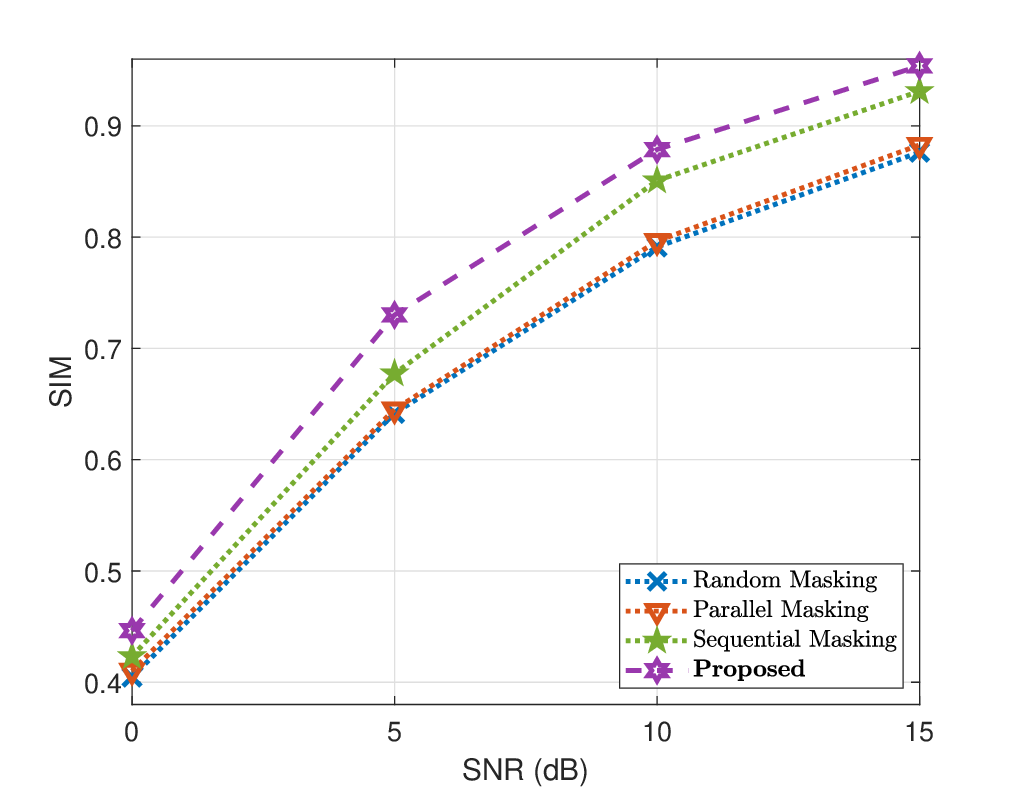,width=8cm}}
    \caption{SIM performance comparison of the joint Tx–Rx strategies with different masking policies.}\vspace{-3mm} 
    \label{Fig: Tx-Rx 1}
\end{figure}

\begin{figure}[t]
    \centering
    \subfigure[The Europarl corpus dataset.]
    {\epsfig{file=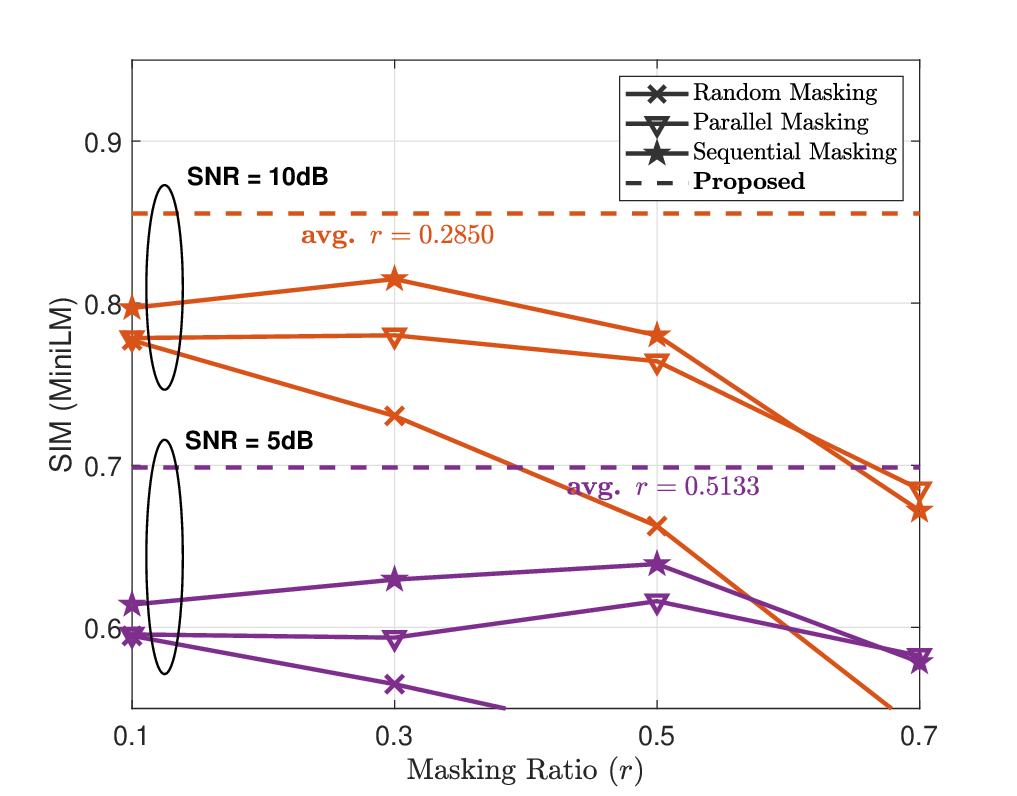,width=8cm}}
    \subfigure[The WikiText-103 dataset.]
    {\epsfig{file=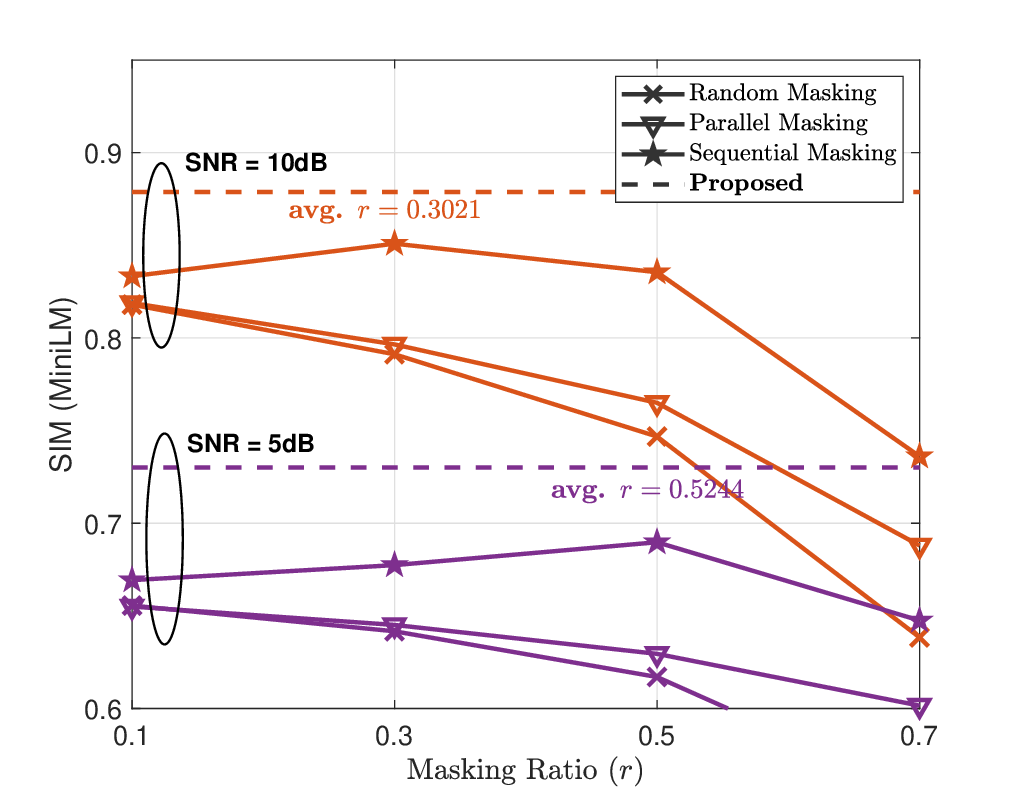,width=8cm}}
    \caption{SIM performance of the joint Tx–Rx strategies versus masking ratio $r$, comparing fixed $r$ masking strategies and the proposed adaptive masking-ratio optimization.}\vspace{-3mm} 
    \label{Fig: Tx-Rx 2}
\end{figure}

Fig. \ref{Fig: Tx-Rx 1} shows the SIM performance of the token communication frameworks with different Tx masking strategies. For the comparison schemes, a fixed masking ratio of $r = 0.3$ is used, whereas the proposed adaptively determines the masking ratio. As shown in Fig. \ref{Fig: Tx-Rx 1}, Sequential Masking consistently achieves higher SIM performance than Random Masking and Parallel Masking for each fixed $r$ across all SNR regimes, as it sequentially refines masking decisions by incorporating the updated contextual priors of masked tokens.
Moreover, among all strategies, the proposed achieves the best SIM performance for both datasets across the entire SNR range. These results confirm the effectiveness of jointly controlling the masking ratio in accordance with the communication environment and token-level context.

Fig. \ref{Fig: Tx-Rx 2} shows the SIM performance of the joint Tx-Rx strategy as a function of the masking ratio $r$.
The Random Masking, Parallel Masking, and Sequential Masking curves represent fixed-ratio schemes, so the x-axis is directly applicable only to these methods. In contrast, the proposed adaptively selects the masking ratio for each channel and token instance and is therefore not tied to a single fixed $r$.
In Fig. \ref{Fig: Tx-Rx 2}, among the fixed-ratio strategies, Sequential Masking consistently attains the highest SIM performance for most values of $r$.
Furthermore, the masking ratio control strategy further improves performance over fixed-ratio masking by adapting $r$ to the channel condition and token instance rather than relying on a preset value.
The reported average optimized ratios are also broadly consistent with the fixed ratios that yield the best Sequential Masking performance. For example, in Fig. \ref{Fig: Tx-Rx 1}(a), the averages $r=0.5133$ and $r=0.2850$ at 5 dB and 10 dB align with the best fixed-ratio points near $r=0.5$ and $r=0.3$, respectively, indicating that the proposed determination effectively identifies appropriate masking levels.

\subsection{Impact of the Rx Iteration Control}

In this subsection, we evaluate the effectiveness of the proposed iteration control strategy introduced in Sec. IV-B. Instead of applying a maximum number of Rx iterations to all tokens, the proposed approach determines the required number of refinement iterations for each token individually based on its posterior uncertainty. As a benchmark, we consider a Maximum Iteration scheme where all tokens are updated up to the maximum number of updates $L_{\rm max}$. 

\begin{figure}[t]
    \centering
    \subfigure[SIM performance.]
    {\epsfig{file=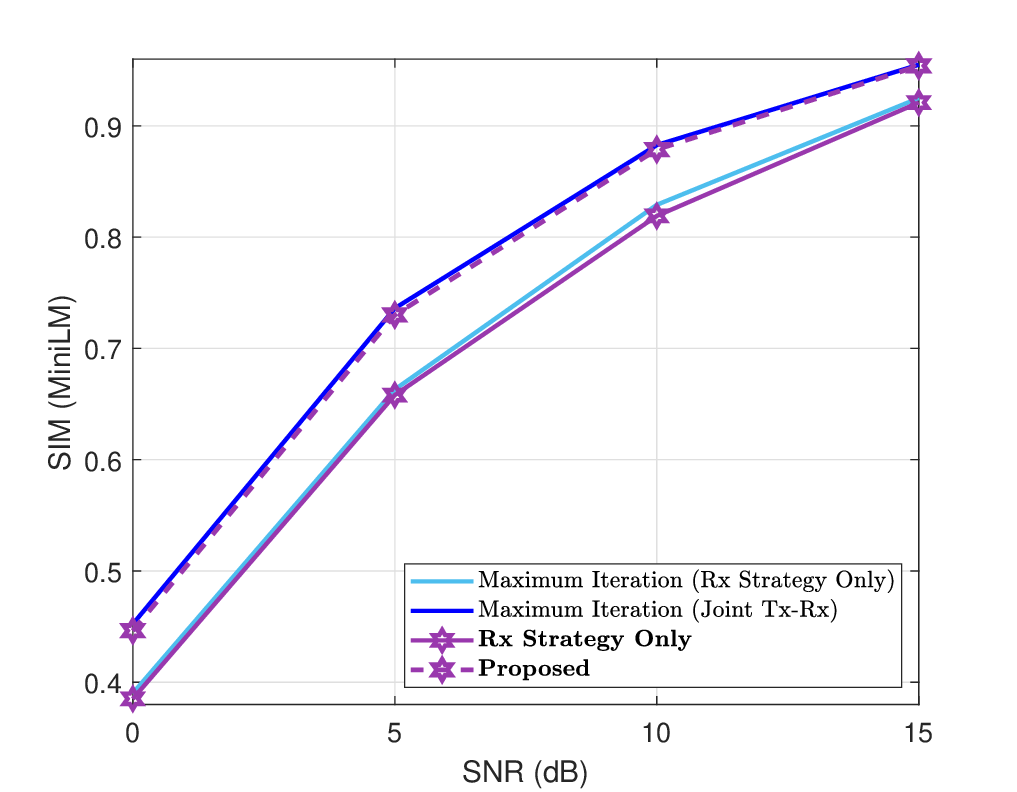,width=8cm}}
    \subfigure[Number of iterations.]
    {\epsfig{file=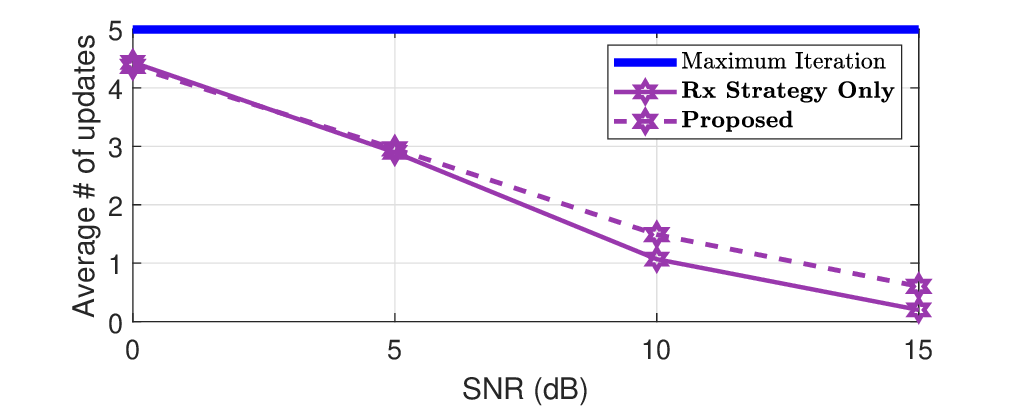,width=8cm}}
    \caption{SIM performance and average number of updates for the joint Tx-Rx strategy with proposed adaptive iteration control using the WikiText-103 dataset.}\vspace{-3mm}
    \label{Fig: Iteration Control}
\end{figure}

Fig. \ref{Fig: Iteration Control} compares the performance of the proposed token-wise iteration determination with this maximum-iteration baseline.
Fig. \ref{Fig: Iteration Control}(a) reports the SIM performance, while Fig. \ref{Fig: Iteration Control}(b) shows the average number of updates per token, defined as $\frac{1}{T}\sum_i L_i$. As shown in Fig. \ref{Fig: Iteration Control}(a), the proposed iteration control achieves SIM performance that is nearly identical to that of the Maximum Iteration baseline across all SNR regimes.
Meanwhile, Fig. \ref{Fig: Iteration Control}(b) shows that the proposed strategy significantly reduces the average number of updates compared to the Maximum Iteration scheme. In particular, as the SNR increases, the number of updates gradually decreases on average, indicating that many tokens can be reliably detected in early iterations without requiring the full $L_{\rm max}$ updates.

\section{Conclusion}

In this paper, we have proposed a context-aware wireless token communication framework that jointly designs token detection at the Rx and token masking at the Tx based on a shared MLM. The Rx performs iterative MAP-based token detection by combining channel observations with MLM-based contextual priors, while the Tx applies context-aware masking to omit highly predictable tokens and concentrate resources on less predictable ones. In addition, the proposed framework proposes key design guidelines, such as the masking ratio control and the detection iteration control, according to channel conditions and token-level contextual priors.
Simulation results have demonstrated that the proposed framework consistently improves the reconstruction quality compared with conventional physical-layer-driven and learning-based baselines.

Future work will extend the proposed framework to more advanced wireless environments, including multiple-input multiple-output (MIMO) and orthogonal frequency division multiplexing (OFDM) systems, where spatial and frequency-domain resources can be jointly optimized with token-level masking strategies. Another promising direction is to generalize the framework to multimodal token communication systems, where contextual priors are formulated across modalities.


\end{document}